\documentclass[letterpaper,english,reprint, aps,prl,superscriptaddress]{revtex4-1}
\usepackage[T1]{fontenc}
\usepackage[latin9]{inputenc}
\usepackage{color}
\usepackage{amsmath}
\usepackage{amssymb}
\usepackage{graphicx}
\usepackage{babel}
\usepackage{todonotes}
\usepackage{verbatim}
\usepackage[normalem]{ulem}

\providecommand{\tabularnewline}{\\}

\begin{document}
\title{Ergodicity Breaking and Scaling Relations for Finite-Time First-Order Phase Transition}
\author{Yu-Xin Wu}
\affiliation{School of Physics, Peking University, Beijing, 100871, China}
\author{Jin-Fu Chen}
\thanks{Corresponding author: jinfuchen@lorentz.leidenuniv.nl}
\affiliation{School of Physics, Peking University, Beijing, 100871, China}
\affiliation{Instituut-Lorentz, Universiteit Leiden, P.O. Box 9506, 2300 RA Leiden, The Netherlands}
\author{H. T. Quan}
\thanks{Corresponding author: htquan@pku.edu.cn}
\affiliation{School of Physics, Peking University, Beijing, 100871, China}
\affiliation{Collaborative Innovation Center of Quantum Matter, Beijing, 100871,
China}
\affiliation{Frontiers Science Center for Nano-Optoelectronics, Peking University,
Beijing, 100871, China}
\date{\today}
\begin{abstract}
Hysteresis and metastable states are typical features associated with ergodicity breaking in the first-order phase transition. We explore the scaling relations of nonequilibrium thermodynamics in finite-time first-order phase transitions. Using the Curie-Weiss model as an example, for large systems we find the excess work scales as $v^{2/3}$ when the magnetic field is quenched at a finite rate $v$ across the phase transition. We further reveal a crossover in the scaling of the excess work from $v^{2/3}$ to $v$ when downsizing the system. Our study elucidates the interplay between the finite-time dynamics and the finite-size effect, which leads to different scaling behaviors of the excess work with or without ergodicity breaking.
\end{abstract}
\maketitle
\textit{\textcolor{magenta}{Introduction}}\textcolor{magenta}{\textendash } 
Phase transitions are drastic changes of a system's  state under variation of the external parameters \citep{Callen_book_1987,Goldenfeld_book_2018}
--for example, the density in a temperature-driven liquid-gas transition, or the magnetization in a paramagnet-ferromagnet transition
\citep{Chaikin_book_1995}.  In the past few decades, owing to the recent development of nonequilibrium thermodynamics
\citep{Graham_PhysRevLett_1984,Keizer_book_1987,Imparato_Elsevier_2007,Seifert_RepProgPhys_2012,Sagawa_JStatMech_2014,Bertini_RevModPhys_2015,Pineda_Entropy_2018,Peliti_book_2021},
there is a growing interest in the field of stochastic thermodynamics with phase transitions \citep{Derrida_JPhysA_1987,Garrahan_PhysRevLett_2007,Mehl_PhysRevE_2008,Lacoste_PhysRevE_2008,Jack_ProgTheorPhysSupp_2010,Ge_JRSoc_2010,Herpich_PhysRevX_2018,Nyawo_PhysRevE_2018,Byrd_PhysRevE_2019,Noa_PhysRevE_2019,Fei_PhysRevLett_2020,Vroylandt_PhysRevLett_2020,Proesmans_PhysicaA_2020,Nguyen_PhysRevE_2020,Meibohm_PhysRevLett_2022,Freitas_PhysRevE_2022}.
A central concept in thermodynamics is the work performed on such a system \citep{Speck_PhysRevE_2004,Speck_EuroPhysJB_2005,Blickle_PhysRevLett_2006,Quan_PhysRevE_2008,Engel_PhysRevE_2009,Nickelsen_EuroPhysJB_2011,Ryabov_JPhysA_2013},
when some external parameters of the system are varied with time. The second law of thermodynamics constrains that the work $W$ performed on the system is no less than the free energy difference $\Delta F$ between the initial and final states. In other words, the excess work done in a finite-rate quench with respect to the quasistatic quench process is non-negative $W-\Delta F\geq0$. A natural question is how thermodynamic quantities such as the excess work scale with
the rate of quench.

In the regular cases without phase transition, it is well-known that in slow isothermal processes, the excess work is proportional to the quench rate as a result of the linear response theory \citep{Salamon_PhysRevA_1980,Salamon_PhysRevLett_1983,Mazonka_Arxiv_1999,Schmiedl_EurophysLett_2007,Crooks_PhysRevLett_2007,Esposito_PhysRevLett_2010,Gong_PhysRevLett_2016,Cavina_PhysRevLett_2017,Ma_PhysRevE_2018,Scandi_Quantum_2019,Ma_PhysRevLett_2020,Chen_PhysRevE_2021}.
While in slow adiabatic processes, the excess work is proportional to the square of the quench rate according to the adiabatic perturbation theory \citep{Sun_JPhysA_1988,Polkovnikov_PhysRevB_2005,Rigolin_PhysRevA_2008,DeGrandi_2010,Polkovnikov_RevModPhys_2011,Chen_PhysRevE_MP_2019}.
In the presence of phase transitions, the scaling behavior becomes much more involved. For the second-order phase transition, it is found that the excess work done during a quench across the critical point exhibits power-law behavior with the quench rate, and the corresponding
exponents are fully determined by the dimension of the system and the critical exponents of the transition \citep{Fei_PhysRevLett_2020,Zhang_PhysRevE_2022},
as in the traditional Kibble-Zurek mechanism \citep{Kibble_JPhysA_1976,Kibble_PhysRep_1980,Zurek_Nature_1985,Zurek_PhysRep_1996,Campo_PhysRevLett_2018,Cui_CommPhys_2020}.
Besides the second-order phase transition, the first-order phase transition is ubiquitous in nature, e.g., in biochemical \citep{Ge_PhysRevLett_2009,Ge_JRSoc_2010,Qian_ChemPhysLett_2016,Pineda_Entropy_2018,Nguyen_PhysRevE_2020},
ecological \citep{Li_PhysicaD_2019}, and electronic systems. Such phase transitions have been studied
in the past three decades \citep{Jung_PhysRevLett_1990,Zhong_1994,hohl_PhysRevLett_1995,Zheng_JPhys_1998,Chakrabarti_1999,Berglund_JPhysA_1999,Zhong_FrontPhys_2016,Scopa_JStatMech_2018,Kundu_PhysRevE_2023}, mainly focusing on dynamical properties in the thermodynamic limit with ergodicity breaking \citep{Palmer1982}.  However, the stochastic thermodynamics of the first-order phase transition remains largely unexplored. Also, 
a central problem in statistical mechanics is how the finite-size effect will influence the thermodynamic properties. For mesoscopic systems, nevertheless, little is known about how the finite-size effect influences the thermodynamic properties of finite-time first-order phase transitions.

In this Letter, we study the ergodicity breaking \citep{Palmer1982} and the scaling behavior of the excess work with the quench rate $v$ in the first-order phase transition. Our focus is put on the interplay between the finite time of the quench process and the finite size of the system. Using the Curie-Weiss model with a varying magnetic field as a paradigmatic example \citep{Fernandez_CommunMathPhys_2012,Collet_JStatPhys_2019,Meibohm_PhysRevLett_2022},
we find the $v^{2/3}$ scaling relation of the excess work with the quench rate for a large system.  
The delay time and the transition time are found to scale as $v^{-1/3}$. When downsizing the system, ergodicity of the system is restored, and the hysteresis shrinks. Meanwhile, the scaling of the excess work transitions from $v^{2/3}$ to $v$. These results will be helpful in optimizing the energy cost and the speed of thermodynamic processes, and may have potential applications in information erasure and heat-engine design.

\textit{\textcolor{magenta}{The model}}\textendash We consider the kinetic Curie-Weiss model with ferromagnetic interaction introduced in Refs. \citep{Meibohm_PhysRevLett_2022,Meibohm_NewJPhys_2023}.
The Curie-Weiss model consists of $N$ Ising spins $\sigma_{i}=\pm1$,
labeled $i=1,\dots,N$, with the coupling strength $J/(2N)$. The system is embedded in a heat reservoir at an inverse temperature $\beta=1/(k_{\mathrm{B}}T)$, and subject to a varying field $H(t)$ controlled by some external agent. The state of the system can be characterized by the magnetic moment
$M=\sum_{i}\sigma_{i}$. A single spin can flip $\pm1\to\mp1$ due to thermal fluctuations of the heat bath, and the magnetic moment changes as $M\to M_{\pm}=M\pm2$. The probability $P(M,t)$ of finding the system in state $M$ at time $t$ evolves under the master equation,
\begin{align}
\frac{\partial P(M,t)}{\partial t} & =\underset{\eta=\pm}{\sum}\Big[W_{\eta}(M_{-\eta},H(t))P(M_{-\eta},t)\nonumber \\
 & \ \ \ \ \ \!\ \ -W_{\eta}(M,H(t))P(M,t)\Big].\label{eq:master eq}
\end{align}
The corresponding transition rates $W_{\pm}(M,H)$ under the external field $H$ are given by \citep{Meibohm_NewJPhys_2023}
\begin{align}
W_{\pm}(M,H)=\frac{N\mp M}{2\tau_{0}}e^{\pm\beta[J(M\pm1)/N+H]},
\label{eq:transition rates}
\end{align}
with microscopic relaxation time $\tau_{0}$. 

In the thermodynamic limit $N\to\infty$, we use the mean magnetization $m\equiv M/N$ to denote the system state. The internal energy density is given by 
\begin{equation}
{\cal H}(m,H)=-\frac{J}{2}m^{2}-Hm.
\end{equation}
The deterministic dynamics is described by the equation of motion of $m(t)$ \citep{SupplMat,Meibohm_PhysRevLett_2022}
\begin{align}
\frac{dm}{dt} & =\frac{2}{\tau_{0}}\Big[\sinh(\beta Jm+\beta H)-m\cosh(\beta Jm+\beta H)\Big]\nonumber \\
 & \equiv f(m,H),\label{eq:equation of motion}
\end{align}
with the microscopic relaxation time $\tau_{0}$.

\begin{figure}
\includegraphics[width=8cm]{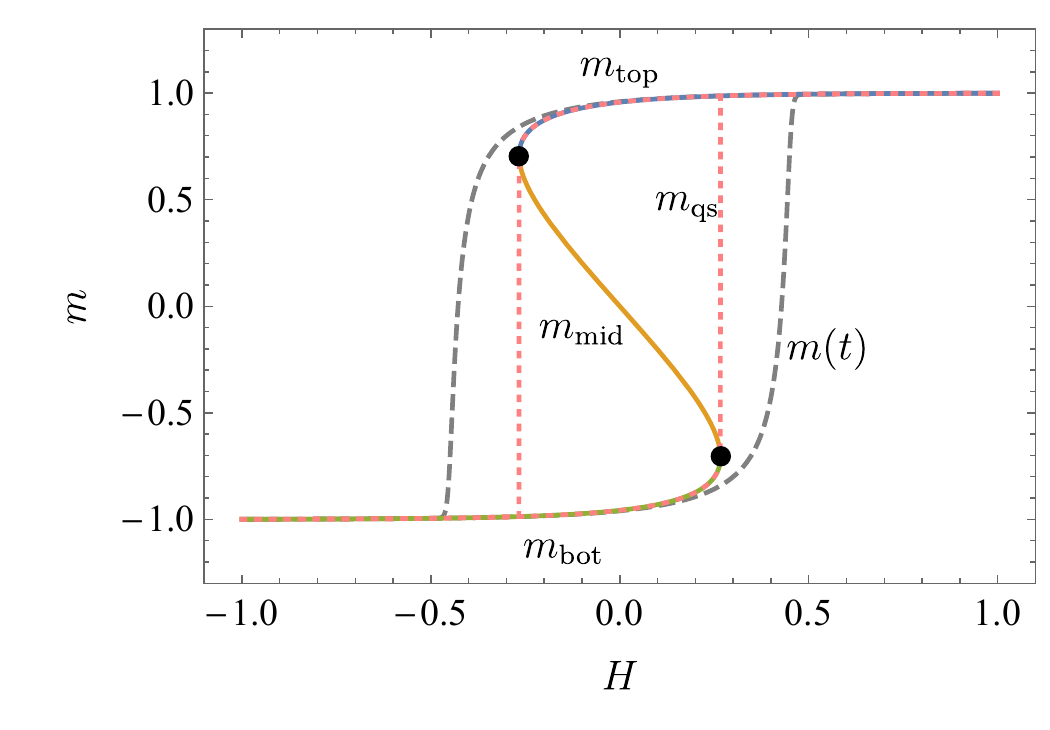} 

\caption{Quasistatic and dynamic hysteresis loops for Curie-Weiss model. The top, middle and bottom solid lines show the steady states $m_{\mathrm{top}},m_{\mathrm{mid}}$, and $m_{\mathrm{bot}}$ as functions of $H$. The black dots denote the right and left turning points $(H_{\pm},m_{\pm})$. The pink dotted lines indicate the quasistatic state $m_{\mathrm{qs}}$. The gray dashed lines represent the transient state of the system $m(t)$ when $H$ is quenched at the rate $v=6$. 
\label{fig:hysteresis}}
\end{figure}

The curve characterizing steady state as a function of $H$ is shown in Fig. \ref{fig:hysteresis}, where we choose the parameters to be $\tau_{0}=2,J=1,\beta=2$. For different $H$, the ordinary
differential equation (\ref{eq:equation of motion}) can have one, two, or three steady-state solutions. For small and large values of $H$, the system has only one steady state, i.e., the system is monostable. For intermediate values of $H$, there are three steady states: $m_{\mathrm{top}}$, $m_{\mathrm{bot}}$, and $m_{\mathrm{mid}}$, i.e., the system exhibits bistability. Among the three, one is globally stable; one is metastable, and the one in between is unstable. The two black dots are the turning points $(H_{\pm},m_{\pm})$, where the metastability disappears.

During the quench process, the external field $H$ is tuned quasistatically from $H_{i}=-1.0$ to $H_{f}=1.0$. The quasistatic hysteresis loop is shown with the pink dotted curve in Fig. \ref{fig:hysteresis}. The quasistatic state $m_{\mathrm{qs}}$ gets trapped in the local minimum as $H$ is increased quasistatically from $H_{i}$ to $H_{f}$, i.e., $m_{\mathrm{qs}}$ follows the lower branch $m_{\mathrm{bot}}$ even when the local minimum is metastable rather than globally stable, until the right turning point $(H_{+},m_{+})$. Then the system state jumps to the upper branch. This is what we refer to as ``quasistatic first-order phase transition.'' So much the same for the return path. The quasistatic state $m_{\mathrm{qs}}$ remains on the upper branch and jumps down to the lower branch at the left turning point $(H_{-},m_{-})$.

When $H(t)$ is tuned at a finite rate $v$, the system tries to keep
pace with \textemdash but ultimately lags behind\textemdash{} the continually changing quasistatic state, as is shown by the gray dashed
curve in Fig. \ref{fig:hysteresis}. This mismatch becomes evident around the turning point. A detailed illustration of the finite-time first-order phase transition can be found in \citep{SupplMat}.

\textit{\textcolor{magenta}{Scaling relations of time}}\textendash During
the quench process, the trajectory follows local equilibrium,
be it metastable or globally stable, until an abrupt transition into
a totally different state from the original one. This transition relies
on the system's history and is regarded as a finite-time first-order phase transition. This is caused by the collapse of the metastable state.
The turning points $(H_{\pm},m_{\pm})=(\pm H^{*},\pm m^{*})$ are solved from
$f(m,H)=0$ and 
$\partial_{m}f(m,H)=0$
as
\begin{align}
m^{*} & =-\sqrt{1-\frac{1}{\beta J}},\label{eq:mb}\\
H^{*} & =\frac{1}{\beta}\Big[\sqrt{\beta J(\beta J-1)}-\mathrm{arctanh}(\sqrt{\frac{\beta J-1}{\beta J}})\Big].
\end{align}

We consider the case in which the external magnetic field $H(t)$
is varied from the initial value $H_{i}$ to the final value $H_{f}$
according to the linear protocol
\begin{equation}
H(t)=H_{i}+(H_{f}-H_{i})\frac{t}{t_{f}},\ \ \ \ \ \ 0\leq t\leq t_f,
\end{equation}
with the quench rate $v=(H_{f}-H_{i})/t_{f}$ and the time duration
$t_{f}$. We label the time when $H(t)$ reaches the right turning
point $H^{*}$ as the turning time $t^{*}$ \citep{Li_PhysicaD_2019}. 

It is important to note that while the potential landscape has drastically
changed from bistable to monostable at the turning time $t^{*}$,
in a finite-rate quench the dynamics of the system cannot catch up
with the change of the potential landscape, i.e., the system displays
a delay in transitioning to the global minimum. The transition occurs
later than the turning time $t^{*}$ by a delay time $\hat{t}_{\mathrm{del}}$
defined through $m(t^{*}+\hat{t}_{\mathrm{del}})=m^{*}$. The delay time depends
on the quench rate $v$.

We further analyze the scaling relation between the delay time and
the quench rate. We shift the variables in the following way: $\hat{H}=H-H^{*},\hat{m}=m-m^{*},\hat{t}=t-t^{*}$.
The evolution equation is turned into
\begin{align}
\frac{d\hat{m}}{d\hat{t}} & =f(m^{*}+\hat{m},H^{*}+v\hat{t})\nonumber\\
 & \approx\frac{2}{\tau_{0}}\Big[\beta J\sqrt{\beta J-1}\hat{m}^{2}+\sqrt{\frac{\beta}{J}}v\hat{t}\Big].\label{eq:approx eq}
\end{align}
In the second line, the equation of motion is expanded around the
turning point for slow quench approximation \citep{SupplMat}. It can be adimensionalized into  \footnote{To demonstrate the generality of the rescaled dynamics in Eq.~\eqref{eq:dimensionless approx eq} for finite-time first-order phase transitions, we derive the same effective equation of motion for both the Kerr model \cite{Drummond_JPhysA_1980,Casteels_PhysRevA_2017,Zhang_PhysRevA_2021} and the Schl$\mathrm{\ddot{o}}$gl model \cite{Schlogl_ZPhysik_1972,Nguyen_PhysRevE_2020,Vellela_JRSoc_2008} in \cite{SupplMat}.}
\begin{equation}
\frac{du}{ds}=u^{2}+s,\label{eq:dimensionless approx eq}
\end{equation}
by rescaling the variables $u=\alpha\hat{m}$,
$s=\gamma\hat{t},$ where
$\alpha=(\beta J\sqrt{\beta J-1})^{2/3}(\tau_{0}v\sqrt{\beta/J}/2)^{-1/3}$
and $\gamma=(\beta J\sqrt{\beta J-1})^{1/3}(4v\sqrt{\beta/J}/\tau_{0}^{2})^{1/3}$.

Under the asymptotic boundary condition $s\to-\infty,u\to\sqrt{-s},$
the solution to Eq. (\ref{eq:dimensionless approx eq}) is the
Airy function \citep{Jung_PhysRevLett_1990}
\begin{equation}
u(s)=\frac{\text{Ai}'(-s)}{\text{Ai}(-s)}.
\end{equation}
In this way, the delay time can be computed as
\begin{align}
\hat{t}_{\mathrm{del}} & =-A_{1}'\bigg[4\sqrt{\beta J(\beta J-1)}\frac{\beta}{\tau_{0}^{2}}v\bigg]^{-1/3},\label{eq:tdel_v analytical}
\end{align}
where $A_{1}'\approx-1.019$ represents the first zero point of the
Airy prime function. 

Besides the delay time $\hat{t}_{\mathrm{del}}$, there is another
time $\hat{t}_{\mathrm{trans}}$ characterizing the non-catching-up
of the dynamics. When the system is quenched at a finite rate, the
transition time $\hat{t}_{\mathrm{trans}}$ to reach the monostable
state, which is defined through $m(t^{*}+\hat{t}_{\mathrm{trans}})=m_{\mathrm{nd}},$
depends on the quench rate $v$. Here $m_{\mathrm{nd}}\approx 1+2e^{-2\beta J-2\sqrt{\beta J(\beta J-1)}}/(2\sqrt{\beta J(\beta J-1)}-2\beta J+1)$ stands for
the nondegenerate root of $f(m,H_{+})=0$.
For $\beta J\gg1$, $m_{\mathrm{nd}}$ can be approximated by $1$. The transition time can be calculated as
\begin{align}
\hat{t}_{\mathrm{trans}} & =-A_{1}\bigg[4\sqrt{\beta J(\beta J-1)}\frac{\beta}{\tau_{0}^{2}}v\bigg]^{-1/3},\label{eq:ttrans_v analytical}
\end{align}
where $A_{1}\approx-2.338$ stands for the first zero point of the
Airy function \citep{Li_PhysicaD_2019}.

\begin{figure}
\includegraphics[width=8cm]{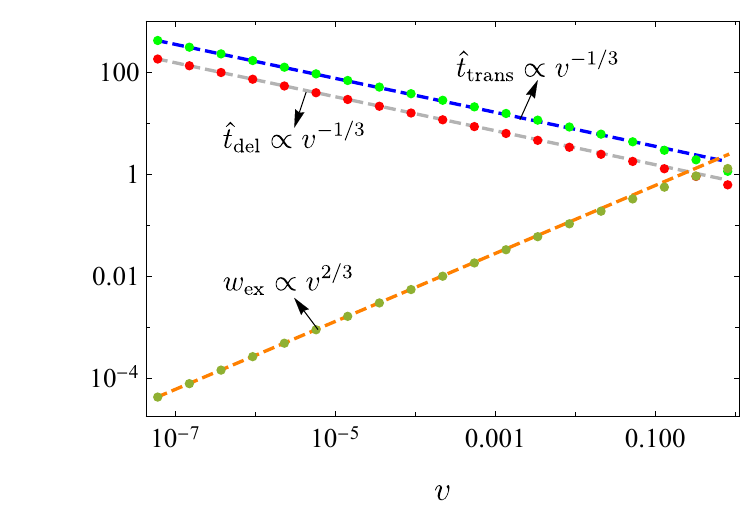}

\caption{The scaling relations of the delay time, the transition time, and the excess work with the quench rate. The delay time $\hat{t}_{\mathrm{del}}$ and the transition time $\hat{t}_{\mathrm{trans}}$
scale as $v^{-1/3}$ while the excess work scales as $v^{2/3}$. The dots are
obtained by numerically solving Eqs. \eqref{eq:equation of motion} and \eqref{eq:trajectory work}. The dashed
lines show the analytical expressions in Eqs. (\ref{eq:tdel_v analytical})
, (\ref{eq:ttrans_v analytical}), and (\ref{eq:Wex_v analytical}). 
\label{fig:scaling relations}}
\end{figure}

Equations (\ref{eq:tdel_v analytical}) and (\ref{eq:ttrans_v analytical})
show that both the delay time and the transition time scale with the
quench rate as $v^{-1/3}$. The matching between analytical expressions
and numerical results for $\hat{t}_{\mathrm{del}}$ and $\hat{t}_{\mathrm{trans}}$
is presented in Fig. \ref{fig:scaling relations}. The exponent
$-$1/3 results from the quadratic leading order $\hat{m}^{2}$ in Eq. \eqref{eq:approx eq} around
the turning point.

\textit{\textcolor{magenta}{Scaling relation of excess work for large systems}}\textendash In a finite-rate quench process, the work performed on the system is greater than that in the quasistatic quench process. In other words, the non-catching-up in the dynamics between the transient state
and the quasistatic state results in the excess work, which characterizes the irreversibility of the process. It is desirable to explore the relation between the excess work and the quench rate.

From the microscopic definition of work \citep{Jarzynski_PhysRevLett_1997,Sekimoto_book_2010},
the work (per site) performed on the system is
\begin{align}
w & =\int_{0}^{t_{f}}dt\frac{\partial{\cal H}}{\partial H}\dot{H}=-v\int_{0}^{t_{f}}dtm(t).\label{eq:trajectory work}
\end{align}
In the quasistatic isothermal process, the work can be calculated from
\begin{equation}
w_{\mathrm{qs}}=\int_{H_{i}}^{H_{f}}dH\frac{\partial{\cal H}(m_{\mathrm{qs}},H)}{\partial H}.
\end{equation}
The excess work, defined as $w_{\mathrm{ex}}=w-w_{\mathrm{qs}},$
describes the excess amount work that one has to perform when the system is quenched at a finite rate rather than quasistatically. In a slow quench, the excess work can be approximated by the enclosed rectangular area between the dynamic and quasistatic hysteresis (see Fig.~\ref{fig:hysteresis})
\begin{align}
w_{\mathrm{ex}} & \approx v\hat{t}_{\mathrm{trans}}(m_{\mathrm{nd}}-m^{*})\nonumber \\
 & \approx\frac{-A_{1}(1+\sqrt{1-\frac{1}{\beta J}})}{\Big[4\sqrt{\beta J(\beta J-1)}\frac{\beta}{\tau_{0}^{2}}\Big]^{1/3}}v^{2/3},\label{eq:Wex_v analytical}
\end{align}
where we approximate $m_{\mathrm{nd}}$ with $1$ in the second line.
Equation (\ref{eq:Wex_v analytical}) shows that the excess work scales with the quench rate as $w_{\mathrm{ex}}\propto v^{2/3}$,
which is verified numerically in Fig. \ref{fig:scaling relations}. This 2/3 scaling relation of the excess work is due to the breaking of ergodicity \citep{Palmer1982} in large systems. 

\textit{\textcolor{magenta}{Scaling relation of excess work for small systems}}\textendash So far, the results above
are caused by neglecting the fluctuations that disappear for infinitely large systems. Nevertheless, for small systems, fluctuations
become non-negligible. A central problem in statistical mechanics is how the finite-size effect will influence the thermodynamic properties. We further study the scaling relation of excess work for small systems. 

\begin{figure}
\includegraphics[width=8.5cm]{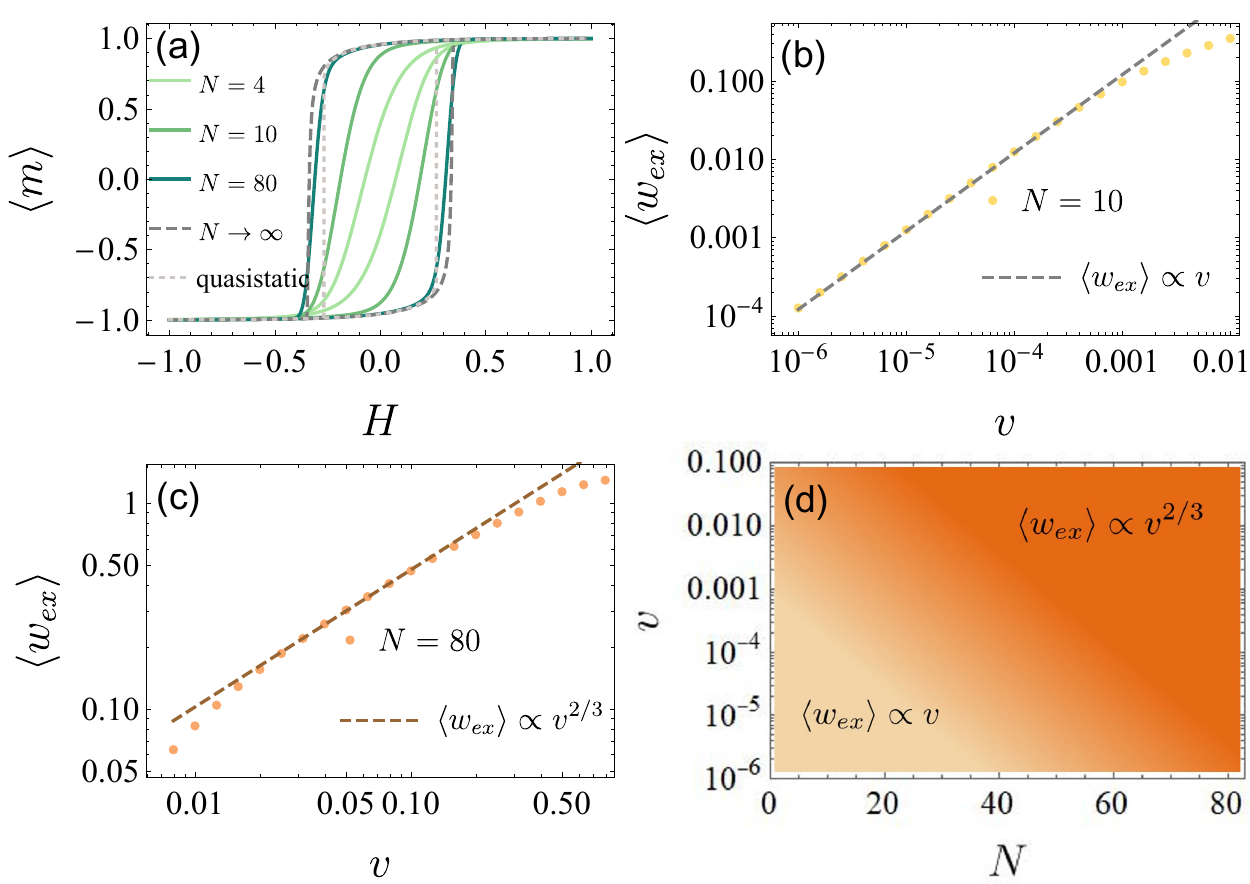}
\caption{The finite-size effect on the dynamic and thermodynamic properties. (a) The hysteresis loop: the average magnetization per site $\langle m \rangle$ as a function of the magnetic field $H$ for
different system sizes $N$. The green solid
lines represent the cases for $N = 4, 10, 80$ from the inside out. The dashed and dotted line, respectively, represent the dynamic and static hysteresis with $N \to \infty$. The quench rate is fixed at the same value $v=0.01$ for all curves. 
(b) For   $N=10$. the average excess work is proportional to the quench rate as $\langle w_\mathrm{ex}\rangle\propto v$.
(c) For $N=80$, the average excess work scales as $\langle w_\mathrm{ex}\rangle\propto v^{2/3}$. The dots in (b) and (c) are obtained from numerically solving Eqs. \eqref{eq:master eq} and \eqref{eq:average work}. 
(d) The crossover between two regimes in the $N-v$ plane. The excess work scales as $\langle w_\mathrm{ex}\rangle\propto v$ in the light-color region, and $\langle w_\mathrm{ex}\rangle\propto v^{2/
3}$ in the dark-color region (see \citep{SupplMat} for details).}
\label{fig:finite_size}
\end{figure}

In order to take fluctuations into consideration,
a stochastic approach has to be envisaged. For finite systems, $M$ is a random variable, and its probability distribution $P(M,t)$ evolves under the master equation (\ref{eq:master eq}). The average magnetization per site $\langle m\rangle=\sum_{M}MP(M,t)/N$
can be calculated from the solution of the master equation \eqref{eq:master eq}. Figure \ref{fig:finite_size}(a) shows the mean magnetization for finite-size systems. The dynamic hysteresis loop approaches the deterministic case (dashed gray curve) as the system size $N$
increases, and shrinks
as $N$ decreases.

The work done on a system during the manipulation of the external
field, along a stochastic trajectory $m(t)$, is given by Eq.
(\ref{eq:trajectory work}). The average work can be calculated from
\begin{equation}
\langle w\rangle=-v\int_{0}^{t_{f}}dt\langle m(t)\rangle.\label{eq:average work}
\end{equation}
The average excess work is defined as the average total work subtracted by
the quasistatic work $\langle w_\mathrm{ex}\rangle=\langle w\rangle-w_\mathrm{qs}.$

For small systems, in the slow driving regime, the quench rate is slower than the system's relaxation rate, which depends on the system size. The system is ergodic and the dynamics falls within the near-equilibrium regime without phase transition. The quasistatic work is equal to the free energy difference that vanishes when $H_{f}=-H_{i}$ (due to the symmetry of Curie-Weiss model).
Thus, the excess work equals the total work, which is proportional
to the quench rate $v$, i.e., $\langle w_\mathrm{ex}\rangle\sim v$ [see
Fig. \ref{fig:finite_size}(b)] as a consequence of the linear response theory \citep{Salamon_PhysRevA_1980,Salamon_PhysRevLett_1983,Mazonka_Arxiv_1999,Schmiedl_EurophysLett_2007,Crooks_PhysRevLett_2007,Esposito_PhysRevLett_2010,Gong_PhysRevLett_2016,Cavina_PhysRevLett_2017,Ma_PhysRevE_2018,Scandi_Quantum_2019,Ma_PhysRevLett_2020,Chen_PhysRevE_2021}. 

For large systems, when the quench rate is slow but not slower than the system's relaxation rate, nonergodic behavior emerges and the quasistatic work can be approximated by that in the deterministic limit. We plot the average excess work $\langle w_\mathrm{ex}\rangle$ as a function of the quench rate
$v$ for system size $N=80$ in Fig. \ref{fig:finite_size} (c). We observe that the average excess work scales as $\langle w_\mathrm{ex}\rangle\propto v^{2/3}$
when the quench rate is slow but not too slow. As the quench further slows down,  ergodicity is gradually restored, leading to the deviation from the 2/3 scaling.

As the system size increases, the scaling of the average excess work
transitions from $\langle w_\mathrm{ex}\rangle\propto v$ to $\langle w_\mathrm{ex}\rangle\propto v^{2/3}$ [see Fig.~\ref{fig:finite_size}(d) and \citep{SupplMat} for details].
This crossover is a consequence of the breaking of ergodicity \citep{Palmer1982}, which has its origin in the interplay between the quench rate and the relaxation rate (system size). For small $N$, thermal fluctuation is strong enough to overcome the energy barrier between two minima. The system is ergodic and the linear scaling relation  $\langle w_\mathrm{ex}\rangle\propto v$ can be explained with the linear response theory. As the system size increases, the first-order phase transition shows up. For large $N$, thermal fluctuation cannot overcome the energy barrier, and ergodicity is broken \citep{Palmer1982}, leading to the $2/3$ scaling of the excess work. Moreover, it is noteworthy that stochastic thermodynamics brings us more information apart from the average value of work. By evaluating the probability distribution of work for finite-size system, Jarzynski's equality is verified \citep{SupplMat}.

\begin{table}[t]
\caption{The scaling relations of the excess work $w_{\mathrm{ex}}$ with the
quench rate $v$ for different situations.\label{tab:comparison}}
\begin{ruledtabular}
\begin{tabular}{lc}
Finite-time isothermal process \citep{Salamon_PhysRevA_1980,Salamon_PhysRevLett_1983,Mazonka_Arxiv_1999,Schmiedl_EurophysLett_2007,Crooks_PhysRevLett_2007,Esposito_PhysRevLett_2010,Gong_PhysRevLett_2016,Cavina_PhysRevLett_2017,Ma_PhysRevE_2018,Scandi_Quantum_2019,Ma_PhysRevLett_2020,Chen_PhysRevE_2021} & $w_{\mathrm{ex}}\propto v$\tabularnewline
Finite-time adiabatic process \citep{Sun_JPhysA_1988,Polkovnikov_PhysRevB_2005,Rigolin_PhysRevA_2008,DeGrandi_2010,Polkovnikov_RevModPhys_2011,Chen_PhysRevE_MP_2019} & $w_{\mathrm{ex}}\propto v^{2}$\tabularnewline
Finite-time first-order phase transition
&  $w_{\mathrm{ex}}\propto v^{2/3}$\tabularnewline
Finite-time second-order phase transition \citep{Fei_PhysRevLett_2020,Zhang_PhysRevE_2022} & $w_{\mathrm{ex}}\propto v^{\delta_{1}}$\tabularnewline
\end{tabular}
\end{ruledtabular}
\end{table}

Last but not least, we would like to compare our results in the first-order
phase transition for open systems with those in the second-order phase transition for isolated quantum systems and the cases
without phase transition in Table. \ref{tab:comparison}. In the second-order phase transition for isolated quantum systems, the  $v^{\delta_{1}}$ scaling behavior of the excess work due to finite-rate quench is obtained for isolated quantum systems in Refs.~\citep{Fei_PhysRevLett_2020,Zhang_PhysRevE_2022}. It is relevant to but different from the Kibble-Zurek scaling for the average density of defects \citep{Kibble_JPhysA_1976,Kibble_PhysRep_1980,Zurek_Nature_1985,Zurek_PhysRep_1996}. The scaling exponent $\delta_{1}$ is determined by the dimension
of the system and the critical exponents of the transition. While the $v^{\delta_{1}}$ scaling behavior is valid for large systems, it transitions \citep{Fei_PhysRevLett_2020} to $v^2$  for small systems without phase transition according to the adiabatic perturbation theorem \citep{Sun_JPhysA_1988,Polkovnikov_PhysRevB_2005,Rigolin_PhysRevA_2008,DeGrandi_2010,Polkovnikov_RevModPhys_2011,Chen_PhysRevE_MP_2019}.
In the first-order phase transition for open systems, we find that the excess work scales as $v^{2/3}$ for large systems where ergodicity is broken \citep{Palmer1982,hohl_PhysRevLett_1995,Zheng_JPhys_1998,Berglund_JPhysA_1999,Zhong_FrontPhys_2016,Scopa_JStatMech_2018,Kundu_PhysRevE_2023}. 
When downsizing the system, there is a crossover in the scaling relation of the excess work from $v^{2/3}$ to $v$. 
In small systems where ergodicity is restored, the scaling relation transitions to $w_{\mathrm{ex}}\propto v$ in finite-time isothermal processes without phase transition, as a result of linear response theory \citep{Salamon_PhysRevA_1980,Salamon_PhysRevLett_1983,Mazonka_Arxiv_1999,Schmiedl_EurophysLett_2007,Crooks_PhysRevLett_2007,Esposito_PhysRevLett_2010,Gong_PhysRevLett_2016,Cavina_PhysRevLett_2017,Ma_PhysRevE_2018,Scandi_Quantum_2019,Ma_PhysRevLett_2020,Chen_PhysRevE_2021}. 

\textit{\textcolor{magenta}{Conclusions}}\textendash In this Letter,
we study the excess work in a finite-time first-order phase transition. We show a $v^{2/3}$ scaling relation of the excess work, which manifests itself as a sign of ergodicity breaking \citep{Palmer1982} in large systems, and its crossover to $v$ when downsizing the system. The underlying mechanism of such a crossover is that ergodicity is restored when decreasing the system size. The hysteresis and metastable state disappear in small systems, and the scaling behavior of the excess work transitions to the case without phase transition. It can be seen that in order to minimize the energy dissipation, one should avoid ergodicity breaking. 
Our results on the crossover in the scaling relation between excess work and quench rate in first-order phase transitions of finite-size systems enhance the understanding of finite-time phase transitions and nonequilibrium thermodynamics in driven systems, with potential applications in information erasure and heat-engine design. Our findings have general relevance
beyond the present model, extending to various systems undergoing the first-order phase transitions \citep{SupplMat}, including open quantum systems \cite{Drummond_JPhysA_1980,Casteels_PhysRevA_2017,Zhang_PhysRevA_2021} and chemical reaction networks \cite{Schlogl_ZPhysik_1972,Nguyen_PhysRevE_2020,Vellela_JRSoc_2008}. In the future, it is desirable to explore the scaling properties of the higher moments of the excess work \citep{Fei_PhysRevLett_2020} and the scaling properties of quantum first-order phase transitions, which would hopefully bring more insights about the nonequilibrium properties of phase transitions.

\begin{acknowledgments}
\textit{Acknowledgements}\textendash -The authors acknowledge helpful discussions with Wei-Mou Zheng. This work is supported by the National Natural Science Foundations
of China (NSFC) under Grants No. 12375028 and No. 11825501, No. 12147157. 
\end{acknowledgments}

\bibliographystyle{apsrev4-1}
\bibliography{scaling_first_order_phase_transition}

\end{document}


\renewcommand{\thefigure}{S\arabic{figure}}
\renewcommand{\theequation}{S\arabic{equation}}
\renewcommand{\bibnumfmt}[1]{[S#1]}
\renewcommand{\citenumfont}[1]{S#1}

\title{Supplemental Meterials of ``Ergodicity Breaking and Scaling Relations for Finite-Time First-Order Phase 
Transition''}
\author{Yu-Xin Wu}
\affiliation{School of Physics, Peking University, Beijing, 100871, China}
\author{Jin-Fu Chen}
\thanks{Corresponding author: jinfuchen@lorentz.leidenuniv.nl}
\affiliation{School of Physics, Peking University, Beijing, 100871, China}
\affiliation{Instituut-Lorentz, Universiteit Leiden, P.O. Box 9506, 2300 RA Leiden, The Netherlands}
\author{H. T. Quan}
\thanks{Corresponding author: htquan@pku.edu.cn}
\affiliation{School of Physics, Peking University, Beijing, 100871, China}
\affiliation{Collaborative Innovation Center of Quantum Matter, Beijing, 100871,
China}
\affiliation{Frontiers Science Center for Nano-Optoelectronics, Peking University,
Beijing, 100871, China}
\date{\today}

\maketitle
We show derivations to the results of the main text. In Section \ref{sec:equation of motion},
we show the derivation of the equation of motion (Eq. (4) in the main
text) from the master equation introduced in Refs. \citep{Meibohm_PhysRevLett_2022, Meibohm_NewJPhys_2023}. Section  \ref{sec:illustration of finite time PT} gives an illustration of the finite-time phase transition. 
In Section \ref{sec:approx equation of motion}, we give the detailed derivation
of the approximate equation of motion (Eq. (8) in the main text)
and its adimensionalization by rescaling the variables.  Section \ref{sec:work distribution} gives the details of the calculation of the work distribution for a finite system. In Section \ref{sec:crossover details}, we show the details of the crossover in the scaling relations. In Section \ref{sec:generality}, we demonstrate the generality of the scaling relations beyond the Curie-Weiss model.

\section{Derivation of the equation of motion\label{sec:equation of motion}}

We derive the equation of motion Eq. (4) in the main text from the
stochastic dynamics introduced in Ref. \citep{Meibohm_PhysRevLett_2022,Meibohm_NewJPhys_2023}.
In the mean-field Curie-Weiss model, the flipping of an arbitrary
spin $\mp1\to\pm1$ leads to the total magnetization to change by
two, i.e., $M\to M_{\pm}\equiv M\pm2$. The evolution of the probability
$P(M,t)$ for finding the system in state $M$ at time $t$ is given
by the master equation
\begin{align}
\frac{dP(M,t)}{dt} & =\underset{\eta=\pm}{\sum}\Big[W_{\eta}(M_{-\eta},H)P(M_{-\eta},t)-W_{\eta}(M,H)P(M,t)\Big].
\end{align}
Here $W_{\pm}(M,H)$ represent the transition rates for $M\to M_{\pm}$ with external field $H$,
which take the form
\begin{align}
W_{\pm}(M,H) & =\frac{N\mp M}{2\tau_{0}}\exp\bigg\{\pm\beta\Big[\frac{J}{N}(M\pm1)+H\Big]\bigg\},\label{eq:transition rates}
\end{align}
with the microscopic relaxation time $\tau_{0}$. 
The transition rates
obey the detailed balance condition 
\begin{align}W_{\pm}(M,H)P^\mathrm{eq}(M,H)=W_{\mp}(M_{\pm},H)P^\mathrm{eq}(M_{\pm},H),\end{align}
where $P^{\mathrm{eq}}(M,H)=\exp[-\beta{\cal F}_{H}(M)]/Z_{H}$ is the equilibrium
distribution with the partition function $Z_{H}$. The free energy takes
the form ${\cal{F}}_{H}(M)=-JM^{2}/(2N)-HM-\beta^{-1}S(M),$ where the internal entropy $S(M)=\ln\Omega(M)$ originates from the microscopic degeneracy $\Omega(M)=N!/\{[(N+M)/2]![(N-M)/2]!\}$.

The ensemble-averaged
magnetization is defined as $\langle M\rangle=\sum_{M}MP(M,t)$. We
obtain the time evolution for the average magnetization as
\begin{align}
\frac{d\langle M\rangle}{dt} & =\sum_M\sum_{\eta=\pm}M\Big[W_{\eta}(M_{-\eta},H)P(M_{-\eta},t)-W_{\eta}(M,H)P(M,t)\Big].
\end{align}

Using the expressions of the transition rates in Eq. (\ref{eq:transition rates}),
the above equation becomes
\begin{align}
\frac{d\langle M\rangle}{dt} & =\underset{M}{\sum}\bigg[\frac{N-M+2}{2\tau_{0}}\exp[\beta J(M-1)/N+\beta H]P(M-2,t)+\frac{N+M+2}{2\tau_{0}}\exp[-\beta J(M+1)/N-\beta H]P(M+2,t)\nonumber \\
 & -\Big(\frac{N-M}{2\tau_{0}}\exp[\beta J(M+1)/N+\beta H]+\frac{N+M}{2\tau_{0}}\exp[-\beta J(M-1)/N-\beta H]\Big)P(M,t)\bigg]M.
\end{align}

In the thermodynamic limit $N\to\infty$, the fluctuation of the magnetization per site $m=M/N$ becomes negligibly small and the system evolves along
a deterministic path $m\approx\left\langle m \right\rangle$. The time evolution
of $\left\langle m\right\rangle$ is
\begin{align}
\frac{d\left\langle m\right\rangle }{dt}&\approx\left\langle \frac{1-M/N}{\tau_{0}}\exp[\beta JM/N+\beta H]-\frac{1+M/N}{\tau_{0}}\exp[-\beta JM/N-\beta H]\right\rangle \\&\approx\frac{1-\left\langle m\right\rangle }{\tau_{0}}\exp[\beta J\left\langle m\right\rangle +\beta H]-\frac{1+\left\langle m\right\rangle }{\tau_{0}}\exp[-\beta J\left\langle m\right\rangle -\beta H],
\end{align}
where in the second equality we use the fact that the distribution of $m$ approaches a delta distribution. In the thermodynamic limit $N\rightarrow\infty$, we can neglect the average symbol $\langle \rangle$ and obtain Eq. (4) in the main text for the deterministic evolution.

\section{Illustration of the finite-time first-order phase transition\label{sec:illustration of finite time PT}}
\begin{figure*}
\includegraphics[width=4.5cm]{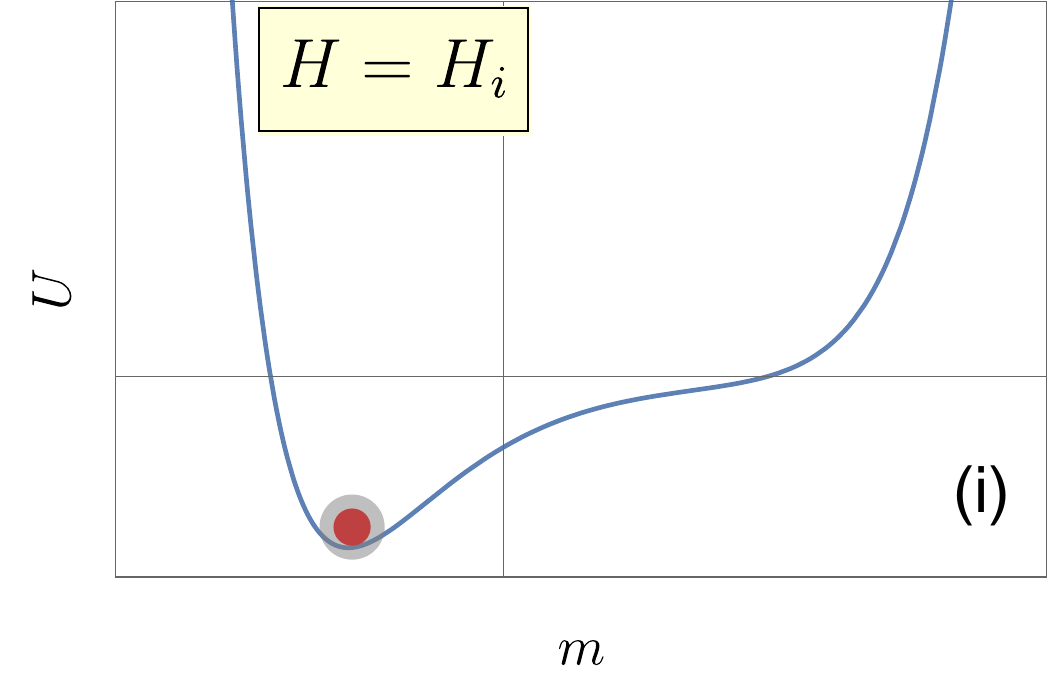}\includegraphics[width=4.5cm]{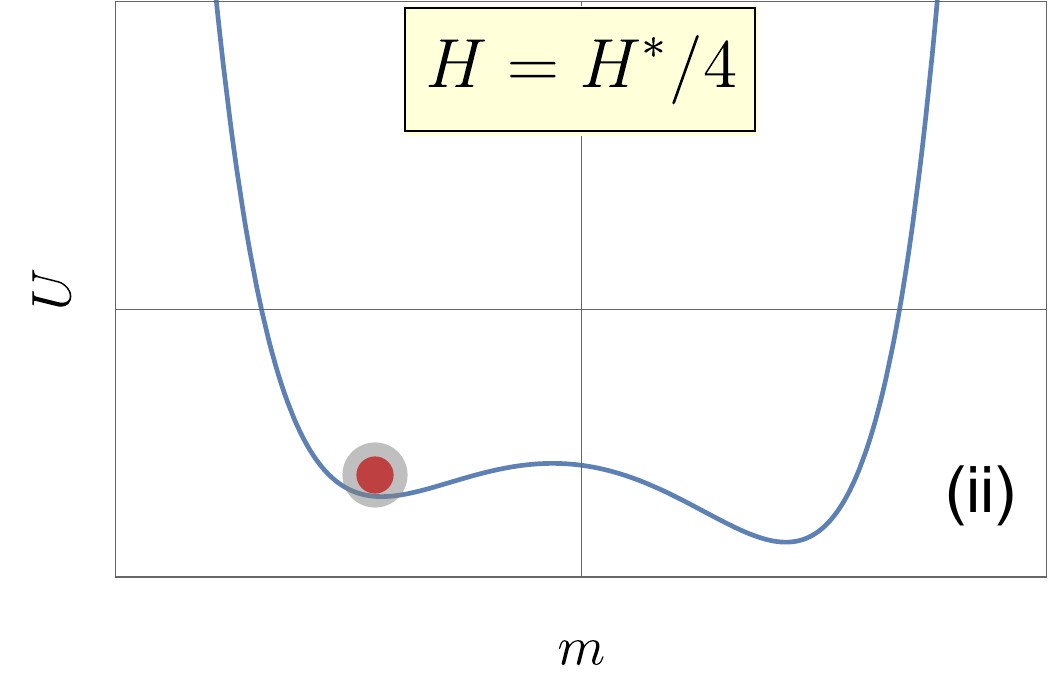}\includegraphics[width=4.5cm]{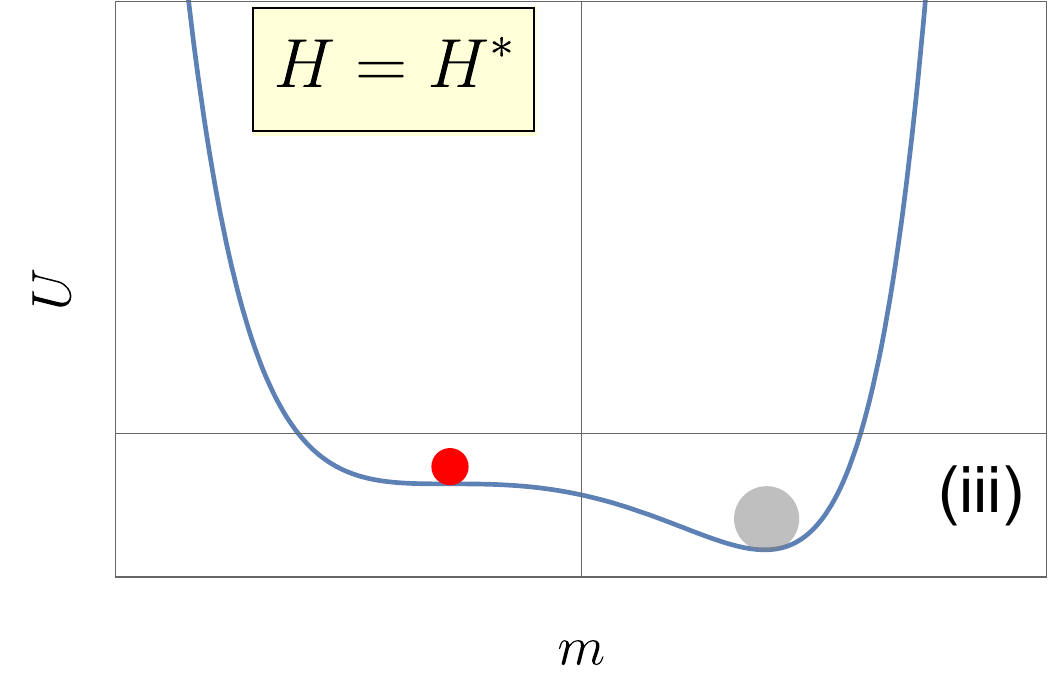}\includegraphics[width=4.5cm]{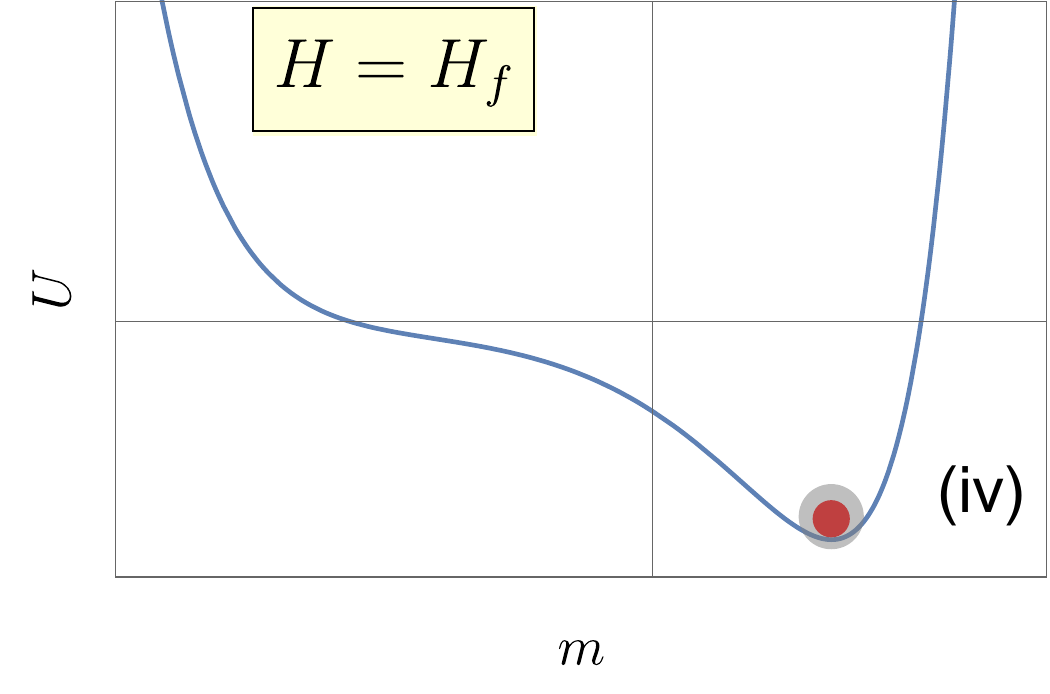}

\caption{The potential landscapes and the evolution of the system. (i-iv) Red
dots show the transient system states $m(t)$ and the gray
circles denote the quasi-static states $m_{\mathrm{qs}}$ for four
$H$ values: $H_{i},\ H^{*}/4,\ H^{*},\ H_{f}$.\label{fig:evolving potential and system state}}
\end{figure*}

We illustrate the finite-time first-order phase transition with a
particle moving in an evolving potential landscape $U$ (where $U$ satisfies $f(m,H)=-\partial_{m}U/\tau_{0}$) in Fig. \ref{fig:evolving potential and system state}.
Far away from the turning point (see Fig. \ref{fig:evolving potential and system state}(i-ii)),
the particle is able to catch up with the change of the potential
landscape and is trapped around the left local minimum, no matter whether it is globally stable or metastable. It becomes different around the turning point (see Fig. \ref{fig:evolving potential and system state}(iii))
where the left local minimum disappears and the system transitions from bistable to monostable. Thus the right minimum becomes the only attractor in the whole landscape. However, the evolution of the particle lags behind the drastic change of the potential. It takes some time for the particle to ``feel the change of its surroundings'' and make a transition to the right stable minimum. The non-catching-up obviously depends on the rate of quench. After reaching the new stable state, the particle gets stuck as the deformation of the potential landscape
continues (Fig. \ref{fig:evolving potential and system state}(iv)).

\section{Derivation of 
 the approximate equation of motion and the adimentionalization\label{sec:approx equation of motion}}

In this section, we show the derivation of approximate equation
of motion and the adimentionalization. We start with Eq. (4) in the main text and shift
the variables to the tuning point as: $\hat{H}=H-H^{*},\hat{m}=m-m^{*},\hat{t}=t-t^{*}$.
The evolution equation is turned into
\begin{align}
\frac{d\hat{m}}{d\hat{t}} & =f(m^{*}+\hat{m},H^{*}+v\hat{t}),\label{eq:translated equation of motion}
\end{align}
where $(H^{*},m^{*})$ is the turning point and $t^{*}$ is the turning
time. For slow quench, we expand the right-hand side to the series
of $\hat{m}$ and $v$ as
\begin{align}
\frac{d\hat{m}}{d\hat{t}} & =\frac{2}{\tau_{0}}\left(\sinh[\beta J(\hat{m}+m^{*})+\beta(v\hat{t}+H^{*})]-(\hat{m}+m^{*})\cosh[\beta J(\hat{m}+m^{*})+\beta(v\hat{t}+H^{*})]\right)\\
 & =\frac{2\sqrt{\beta J}}{\tau_{0}}\sqrt{\left(\hat{m}\sqrt{1-\frac{1}{\beta J}}+\frac{1}{\beta J}\right)^{2}-\hat{m}^{2}}\sinh\left[\beta(J\hat{m}+v\hat{t})-\mathrm{arctanh}(\frac{\hat{m}}{\hat{m}\sqrt{1-\frac{1}{\beta J}}+\frac{1}{\beta J}})\right]\\
 & \approx\frac{2\sqrt{\beta J}}{\tau_{0}}\sqrt{\left(\hat{m}\sqrt{1-\frac{1}{\beta J}}+\frac{1}{\beta J}\right)^{2}-\hat{m}^{2}}\left(\sinh(\beta v\hat{t})+\beta^{2}J^{2}\sqrt{1-\frac{1}{\beta J}}\cosh(\beta v\hat{t})\hat{m}^{2}\right)\\
 & \approx\frac{2}{\tau_{0}}\left(\beta J\sqrt{\beta J-1}\hat{m}^{2}+\sqrt{\frac{\beta}{J}}v\hat{t}\right).
\end{align}
The second line is the precise evolution. The third line is expanded
to the leading order $\hat{m}$ and the fourth line is kept to the
leading order of $v$.

\begin{figure}
\includegraphics[width=8cm]{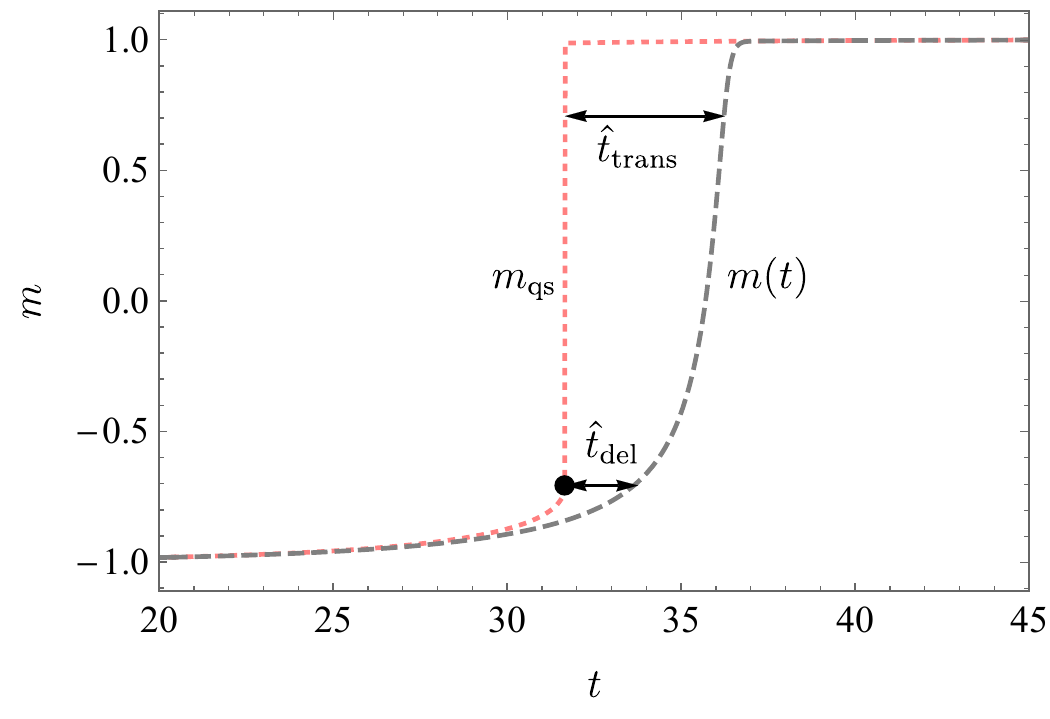}
\caption{The time evolution of the transient state, quasi-static state, the delay time $\hat{t}_{\mathrm{del}}$
and the transition time $\hat{t}_{\mathrm{trans}}$ obtained for finite-rate quench
in the Curie-Weiss model. The black dot denotes the turning point.
The pink dotted line indicates the quasi-static state $m_{\mathrm{qs}}$.
The gray dashed line represents the transient state of the system
$m(t)$ when $H$ is quenched at the rate $v=0.04$. The bottom and the
top double-headed arrows show the delay time $\hat{t}_{\mathrm{del}}$
and the transition time $\hat{t}_{\mathrm{trans}}$, respectively.
\label{fig:mt_tdel_ttrans}}
\end{figure}

Thus we obtain the evolution equation as
\begin{align}
\frac{d\hat{m}}{d\hat{t}} & =\frac{2}{\tau_{0}}\Big[\beta J\sqrt{\beta J-1}\hat{m}^{2}+\sqrt{\frac{\beta}{J}}v\hat{t}\Big]\label{eq:approx eq}\\
 & =A\hat{m}^{2}+B\hat{t},
\end{align}
where we define
\begin{align}
A & =\frac{2}{\tau_{0}}\beta J\sqrt{\beta J-1},\\
B & =\frac{2}{\tau_{0}}\sqrt{\frac{\beta}{J}}v.
\end{align}
By rescaling the variables
\begin{align}
u & =\alpha\hat{m},\\
s & =\gamma\hat{t},
\end{align}
we obtain 
\begin{equation}
\frac{du}{ds}=\frac{A}{\gamma\alpha}u^{2}+\frac{\alpha B}{\gamma^{2}}s.
\end{equation}
The rescaled coefficients are set to unity 
\begin{align}
\frac{A}{\gamma\alpha} & =1,\\
\frac{\alpha B}{\gamma^{2}} & =1,
\end{align}
which gives 
\begin{align}
\alpha & =A^{2/3}B^{-1/3}=\left(\frac{2}{\tau_{0}}\beta J\sqrt{\beta J-1}\right)^{2/3}\left(\frac{2}{\tau_{0}}\sqrt{\frac{\beta}{J}}v\right)^{-1/3},\\
\gamma & =A^{1/3}B^{1/3}=\left(\frac{2}{\tau_{0}}\beta J\sqrt{\beta J-1}\right)^{1/3}\left(\frac{2}{\tau_{0}}\sqrt{\frac{\beta}{J}}v\right)^{1/3}.
\end{align}
Then the differential equation takes the simple form
\begin{equation}
\frac{du}{ds}=u^{2}+s.\label{eq:dimensionless approx}
\end{equation}
This is Eq. (9) in the main text.

\section{Calculation of work distribution via the Feynman-Kac approach\label{sec:work distribution}}

In the main text, the average work is calculated from the probability
distribution of the state. To get more information about the work fluctuation
in finite-size systems, we adopt the Feynman-Kac approach to compute the
distribution of work. 

During the manipulation of the external field $H$ from initial value
$H_{i}$ to $H_{f}$ over the time interval $t_{f}$, the evolution of the system
is described by a stochastic trajectory $M(t)$, and the probability distribution
of the state is $P(M,t)$. The master equation governs the evolution of this distribution 
\begin{equation}
\frac{\partial P(M,t)}{\partial t}=\hat{{\cal L}}P,
\end{equation}
where $\hat{{\cal L}}$ is the evolution matrix. The work done on the
system reads
\begin{equation}
W=\int_{0}^{t_{f}}dt\dot{H}(t)\frac{\partial{\cal F}_{H}(M(t))}{\partial H}.
\end{equation}
The joint probability distribution of finding the system in the state
$M$ and having received work $W$ at time $t$ is denoted as $\Phi(M,W;t)$.
The evolution equation for the joint probability $\Phi(M,W;t)$ can
be derived following the steps of Refs. \citep{Imparato_EurophysLett_2005,Imparato_PhysRevE_2005}:
\begin{equation}
\frac{\partial\Phi}{\partial t}=\hat{{\cal L}}\Phi-\dot{H}\frac{\partial{ {\cal F}_{H}}}{\partial H}\frac{\partial\Phi}{\partial W}.
\end{equation}
The state-dependent generating function  with respect
to the joint probability $\Psi(M,s;t)$ is defined as
\begin{equation}
\Psi(M,s;t)=\int dWe^{isW}\Phi(M,W;t).
\end{equation}
The time evolution of the state-dependent generating function is governed by
the tilted master equation:
\begin{align}
\frac{\partial\Psi}{\partial t}  =&\hat{{\cal L}}\Psi+is\dot{H}\frac{\partial{ {\cal F}_{H}}}{\partial H}\Psi\label{eq:tilted_master_eq}\\
  =&\Big[W_{+}(M-2,H)\Psi(M-2,s;t)+W_{-}(M+2,H)\Psi(M+2,s;t)
 \nonumber\\
 &-\big(W_{+}(M,H)+W_{-}(M,H)-is\dot{H}M\big)\Psi(M,s;t)\Big],\label{eq:tilde master equation}
\end{align}
with the initial condition $\Psi(M,s;0)=P^\mathrm{eq}(M,{H_{i}})=\exp{(-\beta{\cal F}_{H_{i}}(M))}/Z_{H_{i}}$.

The state-independent generating function for work is 
\begin{equation}
\Gamma(s,t)=\int dM\Psi(M,s;t).
\end{equation}
The generating function for work at final time $\Gamma(s,t_{f})$
satisfies Jarzynski's equality \citep{Jarzynski_PhysRevLett_1997}:
\begin{equation}
\langle e^{-\beta W}\rangle=\Gamma(i\beta,t_{f})=\frac{Z_{H_{f}}}{Z_{H_{i}}}.\label{eq:Jarzynski equality}
\end{equation}

In the master equation describing the time evolution of the state
probability distribution, the Liouvillian matrix $\hat{{\cal L}}$
is asymmetric, it can be mapped to a symmetric matrix by introducing
\citep{Matsuo_JStatPhys_1977}
\begin{equation}
\phi(M,t)=\frac{P(M,t)}{\sqrt{P^\mathrm{eq}(M,H(t))}}.
\end{equation}
The differential equation for $\phi(M,t)$ follows
\begin{align}
\frac{d\phi(M,t)}{dt}&=-[\hat{{\cal H}}\phi]_{M}-\frac{\dot{H}}{2}\frac{\partial P^{\mathrm{eq}}(M,H)/\partial H}{P^{\mathrm{eq}}(M,H(t))}\phi(M,t).
\end{align}
The symmetric matrix $\hat{{\cal H}}$ takes the form
\begin{align}
\hat{{\cal H}} & =-\frac{1}{\sqrt{P^\mathrm{eq}}}\hat{{\cal L}}\sqrt{P^\mathrm{eq}}\\
 & =\left(\begin{array}{ccccc}
W_{+}(-N) & -\sqrt{W_{+}(-N)W_{-}(-N+2)} & 0 & 0 & 0\\
-\sqrt{W_{+}(-N)W_{-}(-N+2)} & W_{+}(-N+2)+W_{-}(-N+2) & \dots & \dots & 0\\
0 & \dots & ... & .... & \dots\\
0 & 0 & ... & W_{+}(N-2)+W_{-}(N-2) & -\sqrt{W_{+}(N-2)W_{-}(N)}\\
0 & 0 & 0 & -\sqrt{W_{+}(N-2)W_{-}(N)} & W_{-}(N)
\end{array}\right).
\end{align}
The differential equation for $\phi$ can be written
in a compact form
\begin{equation}
\frac{d\phi}{dt}=-\bigg[\hat{{\cal H}}+\frac{\dot{H}}{2P^\mathrm{eq}}\frac{\partial P^\mathrm{eq}}{\partial H}\bigg]\phi.
\end{equation}
\begin{figure}
\includegraphics[width=8cm]{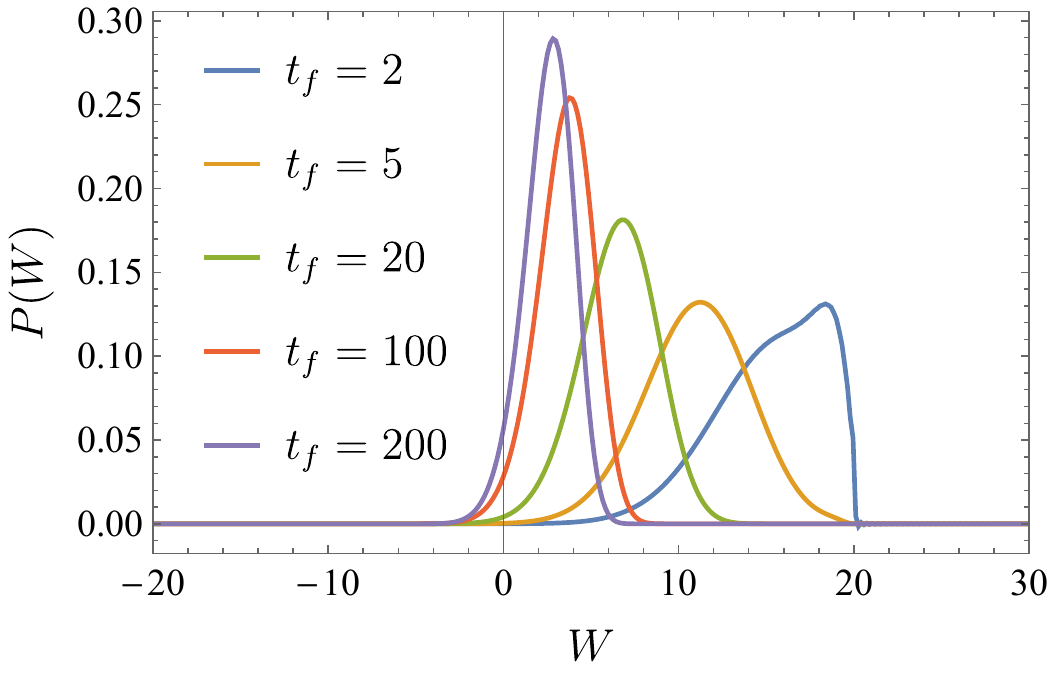}

\caption{The work distribution in the nonequilibrium process when the external field is quenched from $H_{i}=-1$
to $H_{f}=1$ for different tuning time $t_{f}$. From the right to
the left, the five curves represent $t_{f}=2,5,20,100,200$, respectively.
The parameters are $\tau_{0}=2,J=1,\beta=2,N=10$. \label{fig:work_distribution}}
\end{figure}

We continue to rewrite the tilted master equation  \eqref{eq:tilted_master_eq} with the above procedure to compute work statistics. We adopt similar symmetrization techniques to $\Psi$ and introduce
$\psi(M,t;s)$ as
\begin{equation}
\psi(M,t;s)=\frac{\Psi(M,t;s)}{\sqrt{P^\mathrm{eq}(M,H(t))}}.
\end{equation}
The differential equation for $\psi(M,t;s)$ follows
\begin{align}
\frac{d\psi(M,t;s)}{dt}&=-\bigg[\hat{{\cal H}}-is\dot{H}\frac{\partial{\cal  F}_{H}}{\partial H}+\frac{\dot{H}}{2P^{\mathrm{eq}}(M,H(t))}\frac{\partial P^{\mathrm{eq}}(M,H)}{\partial H}\bigg]\psi(M,t;s).
\end{align}
We further introduce a tilted Hamiltonian
\begin{align}
\tilde{\mathcal{H}}(H(t);s) & \equiv\hat{{\cal H}}-i s\dot{H}\frac{\partial{\cal F}_{H}}{\partial H}+\frac{\dot{H}}{2P^\mathrm{eq}}\frac{\partial P^\mathrm{eq}}{\partial H}\\
 & =\hat{{\cal H}}-\frac{\dot{H}}{2}\bigg[(2is+\beta)\frac{\partial{\cal F}_{H}}{\partial H}+\frac{1}{Z_{H}}\frac{\partial Z_{H}}{\partial H}\bigg].
\end{align}
The evolution of $\psi$ is governed by the tilted Hamiltonian
\begin{equation}
\frac{d\psi}{dt}=-\tilde{\mathcal{H}}(H(t);s)\psi,
\end{equation}
with the initial condition
\begin{equation}
\psi(0)=\sqrt{P^\mathrm{eq}(M,0)}.
\end{equation}

At the end of the evolution, the state-independent generating function
of work can be computed by
\begin{equation}
\Gamma(t_{f};s)=\underset{M}{\sum}\sqrt{P^\mathrm{eq}(M,t_{f})}\psi(M,t_{f};s).
\end{equation}
It is related to the work probability distribution by the inverse Fourier
transform
\begin{equation}
P(W,t_{f})=\frac{1}{2\pi}\int dse^{-isW}\Gamma(s,t_{f}).
\end{equation}
We calculate the work distribution of the quench process for finite system sizes of the Curie-Weiss model. The work distribution is plotted in Fig. \ref{fig:work_distribution}.
The results show that as the tuning slows down,  the peak of the work distribution
becomes more and more prominent. The center
approaches the origin since we have chosen $H_{f}=-H_{i}$, leading to the work in the quasi-static
process $W=\Delta F=0$.

\section{Details of the crossover in the scaling relations\label{sec:crossover details}}

In Fig. 3 of the main text, we show a crossover in the scaling relations of the average excess work with the quench rate from $v$ to $v^{2/3}$ as the system size $N$ increases. In the following, we show the details of the crossover in the scaling relations.

\begin{figure}
    \centering
    \includegraphics[width=0.23\linewidth]{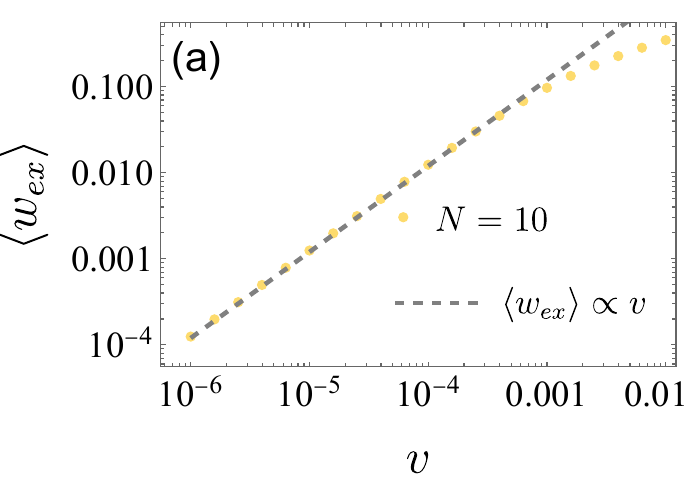}
    \includegraphics[width=0.23\linewidth]{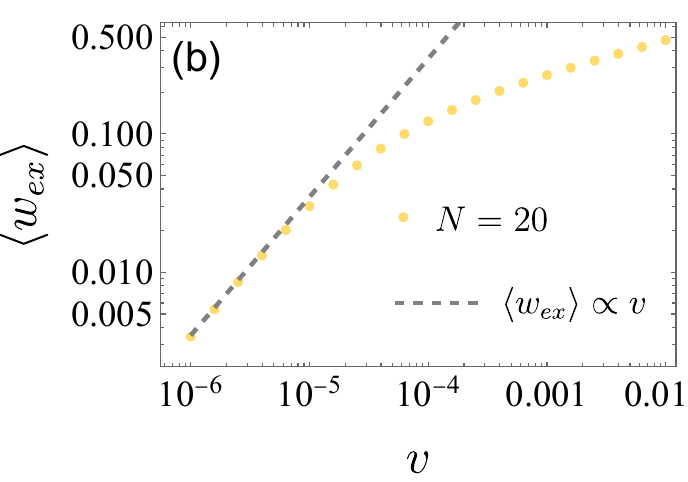}
    \includegraphics[width=0.23\linewidth]{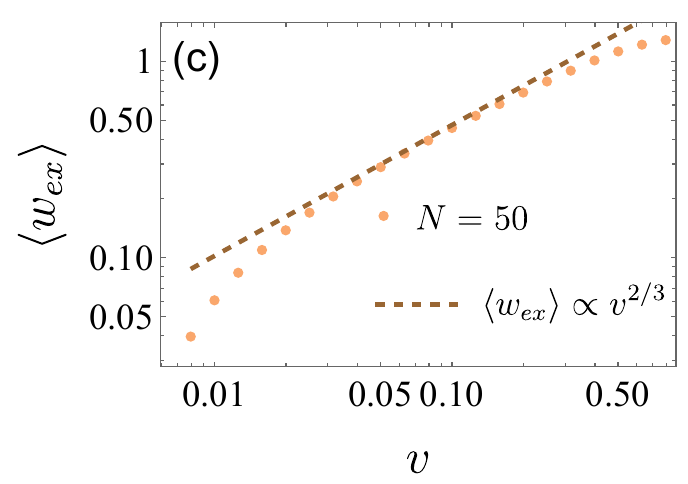}
    \includegraphics[width=0.23\linewidth]{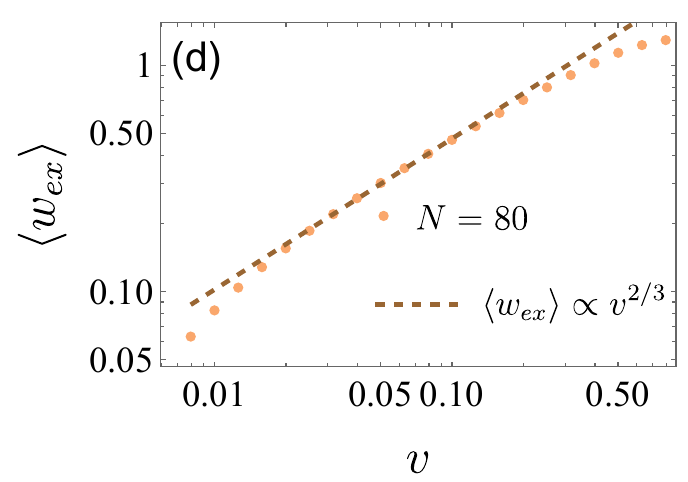}
    \caption{Details of the crossover in the scaling relations of the average excess work with the quench rate. (a)-(d) The average excess work $\langle w_{ex} \rangle$ as a function of the quench rate $v$ for system sizes $N=10,20,50,80$. }
    \label{fig:crossover_details}
\end{figure}

As we can see from Fig. \ref{fig:crossover_details}(a) and (b), the average excess work scales linearly with the quench rate as $\langle w_{ex} \rangle \propto v$ in the slow driving regime when the system size is small. 
This linear relation is caused by the preservation of ergodicity in small systems. 
For smaller systems, the linear scaling holds over a wider range, as can be seen in Fig. \ref{fig:crossover_details}(a) for $N=10$, where more numerical dots fit the linear scaling compared to Fig. \ref{fig:crossover_details}(b) for $N=20$. 
As the system size $N$ increases, ergodicity is gradually broken, leading to the crossover of the scaling relation from $\langle w_{ex} \rangle \propto v$ to $\langle w_{ex} \rangle \propto v^{2/3}$. 
In large systems, e.g., $N=50,80$ in Fig. \ref{fig:crossover_details}(c) and (d), the average excess work scales as $v^{2/3}$ when the quench rate is small but not too small. 
The validity range of the $v^{2/3}$ scaling expands as $N$ increases. It can be seen that more points align with the $v^{2/3}$ line in Fig. \ref{fig:crossover_details}(d) for $N=80$ in comparison with Fig. \ref{fig:crossover_details}(c) for $N=50$.

\section{\label{sec:generality}Generality of the scaling relations}

It is noteworthy that our results about the scaling relations of the excess work are not restricted to the Curie-Weiss model, but are general for systems
undergoing the first-order phase transition with the detailed balance condition. It originates from the time scaling (as can be seen from Eqs. (12) and (15) in the main text), which is universal for systems undergoing the first-order phase transition with and without the detailed balance condition. In the following, we demonstrate that the time scaling $\hat{t}_{\mathrm{trans}} \propto v^{-1/3}$ has relevance beyond the Curie-Weiss model by formulating a general theory and providing two additional examples: the Kerr model and the Schl$\mathrm{\ddot{o}}$gl model.

\subsection{\label{sec:general setup}General setup and adiabatic elimination}

If the order parameter is one-dimensional, the derivation has been given in the main text. Examples includes the Curie-Weiss model and the Schl$\mathrm{\ddot{o}}$gl model (see Sec.~\ref{sec:Schlogel model} below). Generally, if the order parameter is more than one dimensional (see the Kerr model in Sec. \ref{sec:kerrmodel} below), the derivation is a bit more tricky. In the following, we give the detailed derivation of a general setup. The dissipative dynamics with the first-order phase transition can be described by 
\begin{equation}
    \frac{d\Gamma}{dt}=f(\Gamma ,H).
\end{equation}
For example, the time-dependent Ginzburg-Landau equation in Section 8.3 of Ref. \cite{Goldenfeld_book_2018}. 
Here the order parameter $\Gamma$ can be a vector effectively parameterizing the nonequilibrium dynamics; $f$ can be a nonlinear vector function, and $H$ corresponds to the external field (the protocol). 

At the turning point, $f(\Gamma^*,H^*)=0$. By shifting the variables as: $\hat{\Gamma}=\Gamma-\Gamma^*,\hat{H}=H-H^*,\hat{t}=t-t^*$, we expand the equation of motion as 
\begin{align}
    \frac{d\hat{\Gamma}}{d\hat{t}}&=f(\Gamma^{*}+\hat{\Gamma},H^{*}+v\hat{t})\\&\approx f(\Gamma^{*},H^{*})+\frac{\partial f}{\partial\Gamma}(\Gamma^{*},H^{*})\hat{\Gamma}+\frac{1}{2}\frac{\partial^{2}f}{\partial\Gamma^{2}}(\Gamma^{*},H^{*})\hat{\Gamma}^{2}+\frac{\partial f}{\partial H}(\Gamma^{*},H^{*})\hat{H}.
\end{align}
In general, the protocol can be nonlinear, but only the quench rate $v$ at the turning point is essential. So we consider a linear quench of $\hat{H}=v\hat{t}$ for simplicity.
The equation of motion for each component $\hat\Gamma_n$ around the turning point can be written as
\begin{align}
    \frac{d}{d\hat{t}}\hat\Gamma_n=\sum_i J_{ni}\hat\Gamma_i+\frac{1}{2}\sum_{ij}J_{nij}\hat{\Gamma}_i\hat{\Gamma}_j+k_n v \hat{t},
\end{align}
where $k_{n}=(\partial f_{n}/\partial H)(\Gamma^{*},H^{*})$, and $J_{ni}$ and $J_{nij}$ represent the first-order and second-order derivatives, respectively.

Due to the first-order phase transition, the unstable property requires that the matrix $J_{ni}$ has some null space. We can assume it is one-dimensional and represented by $\hat\Gamma_0$ (Otherwise one can find the null space of $J_{ni}$ and do a similarity transform on $\hat{\Gamma}$). The differential equation for this unstable mode is 
\begin{align}
    \frac{d}{d\hat{t}}\hat{\Gamma}_{0}=\frac{1}{2}\sum_{ij}J_{0ij}\hat{\Gamma}_{i}\hat{\Gamma}_{j}+k_{0}v \hat{t}.\label{eq:dGammahat0dt}
\end{align}
Since other modes are stable within a finite relaxation time, it is sufficient to keep the first-order term
\begin{align}
    \frac{d}{d\hat{t}}\hat{\Gamma}_{n}=\sum_{i}J_{ni}\hat{\Gamma}_{i}+k_{n}v \hat{t}.
\end{align}
When the quench is very slow $v\hat{t}\rightarrow 0$, the relaxation of other modes is fast enough to apply the adiabatic elimination such that 
\begin{align}  0=\sum_{i}J_{ni}\hat{\Gamma}_{i}+k_{n}v \hat{t}.
\end{align}
For dissipative dynamics, all other modes have finite dissipative timescales, which ensures the unique solution $\hat{\Gamma}_i$ as a function of $\hat{\Gamma}_0$.
Namely, they can converge to steady-state values according to the slow driving assumption.

The only relevant term in $J_{0ij}\hat{\Gamma}_i\hat{\Gamma}_j$ of Eq.~\eqref{eq:dGammahat0dt} is proportional to $\hat{\Gamma}_0^2 $, since the other terms are either proportional to $\hat{\Gamma}_0vt$ or $v^2\hat{t}^2$.
Keeping only the relevant terms, we obtain 
\begin{align}
\frac{d}{d\hat{t}}\hat{\Gamma}_{0}&=\frac{1}{2}J_{0}\hat{\Gamma}_{0}^{2}+k_{0}v\hat{t},
\label{eq:general_EMO}
\end{align}
where the effective $J_0$ can be solved from the adiabatic elimination procedure. 
From Sec. III of the Supplementary Materials we know an equation of motion such as Eq. (\ref{eq:general_EMO}) with the quadratic leading order around the turning point always gives rise to the $-1/3$ scaling of time (see Eqs. (11) and (12) in the main text).
We further apply this framework to derive the effective dynamics in specific models. 

\subsection{Kerr model}\label{sec:kerrmodel}
We consider a Kerr oscillator subject to two-photon driving and single-photon loss. It is a model studied in open quantum systems and quantum optics that exhibits a first-order phase transition and hysteresis \cite{Drummond_JPhysA_1980,Casteels_PhysRevA_2017,Zhang_PhysRevA_2021}. However, its nonequilibrium dynamics during the finite-time driving remains largely unexplored. 

The density matrix $\rho$ of the system evolves according to the Lindblad master equation 
\begin{align}
    \frac{d}{dt}\rho=-i[H,\rho]+\gamma (a \rho a^\dagger -\frac{1}{2}\{a^\dagger a,\rho\}).
\end{align}
where $a^\dagger$ and $a$ are the creation and annihilation operators of a photon mode and $\gamma$ characterizes the rate of the single-photon loss. The Hamiltonian is 
\begin{align}
    H=-\omega a^\dagger a+\frac{U}{2} a^\dagger a^\dagger a a +\frac{G}{4} (a^\dagger a^\dagger +a a),
\end{align}
where $\omega$ represents the detuning between the frequency of the cavity and the driving; $U$ characterizes the nonlinear potential; $G$ is the strength of the two-photon driving.

Setting $\alpha=\langle a\rangle$ and $\alpha^*=\langle a^\dagger\rangle$, the semi-classical approximation gives the differential equation for the expectation value
\begin{align}
    \dot{\alpha}=(i \omega-i U|\alpha|^2-\frac{\gamma}{2})\alpha-i\frac{G}{2}\alpha^*,
\end{align}
where the real part $x=\mathrm{Re}(\alpha)$  and imaginary part $y=\mathrm{Im}(\alpha)$ satisfy
\begin{align}
\dot{x}&=-[\omega-U(x^{2}+y^{2})]y-\frac{\gamma}{2}x-\frac{G}{2}y,\\\dot{y}&=[\omega-U(x^{2}+y^{2})]x-\frac{\gamma}{2}y-\frac{G}{2}x.
\end{align}
When $G<\gamma$, the average photon number becomes $n=|\alpha|^2=0$; when $\gamma\leq G\leq\sqrt{\gamma^2 +4 \omega^2 }$, $n=0$ or $\omega+\sqrt{|G|^2-\gamma^2}/(2U)$; when $G>\sqrt{\gamma^2 +4 \omega^2 }$, $n=\omega+\sqrt{|G|^2-\gamma^2}/(2U)$.
We consider $G=\gamma-v\hat{t}$ which decreases from a large value to $G=0$. 
The finite-time first order phase transition occurs at the turning point $G=\gamma$, where there are three steady-state solutions $x=0,y=0$, $x=-\sqrt{\omega/(2U)},y=\sqrt{\omega/(2U)}$, and $x=\sqrt{\omega/(2U)},y=-\sqrt{\omega/(2U)}$. The last two become unstable when $G\leq\gamma$.
We shift the variables to the turning point as $\hat{x}=x+\sqrt{\omega/(2U)}$ and $\hat{y}=y-\sqrt{\omega/(2U)}$.

To analyze the critical behavior, we expand the differential equations around the turning point and retain only the second-order terms
\begin{align}
    \frac{d}{d\hat{t}}\hat{x}&=-\left(\omega+\frac{\gamma}{2}\right)\hat{x}+\left(\omega-\frac{\gamma}{2}\right)\hat{y}+\sqrt{\frac{\omega U}{2}}\hat{x}^{2}+3\sqrt{\frac{\omega U}{2}}\hat{y}^{2}-\sqrt{2\omega U}\hat{x}\hat{y}+\sqrt{\frac{\omega}{2U}}\frac{v\hat{t}}{2},\\\frac{d}{d\hat{t}}\hat{y}&=-\left(\omega+\frac{\gamma}{2}\right)\hat{x}+\left(\omega-\frac{\gamma}{2}\right)\hat{y}+3\sqrt{\frac{\omega U}{2}}\hat{x}^{2}+\sqrt{\frac{\omega U}{2}}\hat{y}^{2}-\sqrt{2\omega U}\hat{x}\hat{y}-\sqrt{\frac{\omega}{2U}}\frac{v\hat{t}}{2}.
\end{align}
The slow dynamics is reflected by $\eta=\hat{x}-\hat{y}$, and we also introduce $\xi=\hat x+\hat y$. The differential equations become
\begin{align}
\frac{d\xi}{d\hat{t}}&=-\gamma\xi-2\omega\eta+3\sqrt{\frac{\omega U}{2}}\eta^{2}+\sqrt{\frac{\omega U}{2}}\xi^{2},\\\frac{d\eta}{d\hat{t}}&=-\sqrt{2\omega U}\xi\eta+\sqrt{\frac{\omega}{2U}}v\hat{t}.
\end{align}

According to our general theory in Sec. \ref{sec:general setup}, we can write the differential equations with the relevant terms as 
\begin{align}
   \frac{d\xi }{d\hat t}&=-\gamma\xi-2\omega\eta,\label{eq:dotxifast}\\\frac{d\eta }{d\hat t}&=-\sqrt{2\omega U}\xi\eta+\sqrt{\frac{\omega}{2U}}v \hat t.
\end{align}
We eliminate the fast dynamics by Eq.~\eqref{eq:dotxifast} adiabatically, i.e., $0=-\gamma \xi-2\omega \eta$, and obtain\begin{align}
    \frac{d\eta }{d\hat t}&=\frac{2\sqrt{2\omega^{3}U}}{\gamma}\eta^{2}+\sqrt{\frac{\omega}{2U}}v\hat t,
\end{align}
which is in the form of Eq. (\ref{eq:general_EMO}). From this, one can obtain the $-1/3$ time scaling.

\subsection{Schl$\mathrm{\ddot{o}}$gl model} \label{sec:Schlogel model}

The Schl$\mathrm{\ddot{o}}$gl model is an autocatalytic, trimolecular reaction scheme widely used to study nonequilibrium first- and second-order phase transitions \cite{Schlogl_ZPhysik_1972,Nguyen_PhysRevE_2020,Vellela_JRSoc_2008}. The model consists of a dynamical chemical species $X$ in a volume $V$ at constant temperature. The external reservoir contains two chemical species $A$ and $B$. The set of chemical reactions is 
\begin{align}
A+2X&\ce{<=>[\textit{k}_1][\textit{k}_2]}3X,\\
B&\ce{<=>[\textit{k}_3][\textit{k}_4]}X,
\end{align}
where $k_{1},k_{2},k_{3},k_{4}$ are reaction rates. 

In the thermodynamic limit $V \to \infty$, the system state is denoted by the concentration of $X$ molecule $x=N_X/V$, where $N_X$ is the number of $X$ molecules. The time evolution of $x$ follows 
\begin{align}
\frac{dx}{dt}
=-k_{2}x^{3} + k_{1}ax^{2} - k_{4}x + k_{3}b(t) .
\end{align}
where $a$ and $b$ are the concentrations of species $A$ and $B$, respectively. The system can undergo a first-order phase transition when $b$ is varied with time and the other parameters are fixed. The turning points $(x_{\pm},b_{\pm})$ are solved as
\begin{align}
    x_{\pm}&=\frac{a k_{1}\mp \sqrt{a^2 k_{1}^2-3 k_{2} k_{4}}}{3 k_{2}},
    \nonumber\\
    b_{\pm}&=\frac{-2 a^3 k_{1}^3+9 a k_{1} k_{2} k_{4}\pm 2  \sqrt{(a^2 k_{1}^2-3 k_{2} k_{4})^3}}{27 k_{2}^2k_{3}}.
\end{align}
Without loss of generality, we assume the protocol is quenched linearly $b=b_{+}+v\hat{t}$ at the rate $v$. The time when the value of $b$ reaches the right turning point $b_{+}$ is denoted as $t_{+}$. By shifting the variables to the turning point $\hat{x}=x-x_{+},\hat{t}=t-t_{+}$, and keeping the leading order term, we obtain
\begin{equation}
    \frac{d\hat{x}}{d\hat{t}} = (-3k_{2}x_{+} +  k_{1}a)\hat{x}^2 + v \hat{t},
\end{equation}
which is in the form of Eq. (\ref{eq:general_EMO}). From this, one can obtain the $-1/3$ time scaling.

\bibliographystyle{apsrev4-1}
\bibliography{scaling_first_order_phase_transition}